# A Comparative Evaluation of Power Converter Circuits to Increase the Power Transfer Capability of High Voltage Transmission Lines

Rafael Castillo-Sierra[a],[b], Giri Venkataramanan[a]

[a] Department of Electrical and Computer Engineering, University of Wisconsin – Madison, Madison, USA
[b] Department of Electrical and Electronic Engineering, Universidad del Norte, Barranquilla, Colombia
castillosier@wisc.edu, giri@engr.wisc.edu

*Abstract*— AC transmission lines with lengths greater than 80km cannot be used to their maximum capacity due to limits in voltage drops and transient stability. This inefficient way of using conductors in a transmission line can be overcome if the electrical frequency at which energy is transmitted is reduced. This is why this work focuses on the comparison of the Modular Multilevel Matrix Converter (MMMC) and the Back-to-Back Modular Multilevel Converter (BTB-MMC), topologies that have shown qualities as frequency converters. For comparison, an analytical model of each topology is used to relate design considerations to their operational variables. Among the aspects to be compared are, the requirements of the semiconductors, the required reactive components, operating losses and fault tolerance. Detailed design equations, EMTP simulations, and comparison table are presented.

## I. INTRODUCTION

Voltage drops and transient stability are inherent problems in transmission lines that force them to be operated below their thermal capacity in order to maintain safe operation within the margins established by the regulations. Both problems are associated with the length of the lines and this in turn affects a crucial parameter such as inductive reactance. The greater this reactance, the greater the voltage drops and the greater the transmission angle, so it is preferred to decrease its transmission capacity in order to keep both parameters within acceptable limits. By reducing the reactance, it is possible to increase the transmission capacity of the lines. In this regard, solutions have been raised in the literature. For example, the use of series compensation with the line to reduce the net reactance [1] or increase the conductor cross section [2]. However, for several years now, a new approach to solving this problem has been proposed. The solution aims to reduce the reactance of the line through the reduction of the electrical frequency. Several works have studied the feasibility of this solution giving favorable conclusions about its practical implementation [3]–[6]. As *Figure 1a* shows, the general idea behind this approach suggests connecting two frequency converters at the ends of the line, which transform the grid frequency ($f_G$) to the transmission frequency ($f_T$) and vice versa. Historically, frequency conversion has been carried out through SCR-based cycloconverters. However, their high harmonic content in these applications makes them unattractive for high voltage applications. Consequently, Modular Multilevel Converter (MMC) [7] and Modular Multilevel Matrix Converter (MMMC) [8] have been introduced as new alternatives in the required frequency conversion. Much of existing literature aim to introduce the principles and establish performance levels, and do not focus on the evaluation of design of the power converter circuit. In order to know the general performance of these new circuits as frequency converters, this work aims to carry out a comparative evaluation of them, considering aspects as semiconductor requirements, reactive component requirement, operating losses and fault tolerance.

The rest of the article is organized as follows, Section II shows the main characteristics of the power circuits considered, in Section III the comparison is made based on the aforementioned criteria and finally conclusions close the paper.

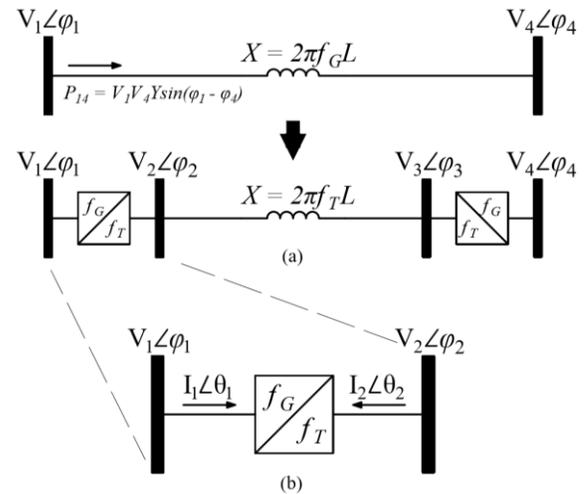

Figure 1. Conceptual schematic, transmission line with frequency converter

## II. REVIEW OF FREQUENCY CONVERTER TOPOLOGIES

In this section, each of the converter topologies is presented, giving a general description of its characteristics and mode of operation. Details regarding component dimensions and overall operation are compared in detail in section III.

### A. Modular Multilevel Matrix Converter (MMMC)

The block shown in *Figure 1b.* can contain an MMMC, which is an electronic converter capable of performing frequency conversion directly, without going through an intermediate DC stage. As shown in *Figure 2a*, the MMMC is made up of nine branches which, through a matrix structure, interconnect the terminals of the input system (*a1, b1, c1*) with the output system (*a2, b2, c2*). Each branch is made up of a controlled voltage source ($v_B$) in series with an inductor ($L_B$). The branch inductor $L_B$ has the functions of (i) avoiding the parallel connection of the internal voltages in order to reduce circulating currents between the branches of the converter [9]–[11], (ii) reduce any high frequency component in the branch current ($i_B$) produced by the switching of the switches, and (iii) serve as a separation between the internal voltages and the voltages of the external networks to the converter for the purpose of controlling the flow of active and reactive power [12]. The controlled voltage source $v_B$, which will be referred to as internal voltage from now on, consists of a string of N submodules connected in series. Each of the submodules corresponds to a Full-Bridge as shown in *Figure 2b*. Such a bridge is made up of fully controlled switches accompanied by an antiparallel diode, allowing bidirectional current flow [13]. The synthesis of the internal voltages is carried out through the sum of the individual voltages of the N submodules in conjunction with a modulation process (e. g. PWM) of the DC voltage of the storage capacitor $C_s$ in each submodule, as shown in (1).

$$v_B(t) = d_B(t) \times v_s^\Sigma(t) \quad (1)$$

Where $v_s^\Sigma$ corresponds to the sum of each of the voltages of the capacitors $C_s$ of the N submodules and $d_B$ corresponds to the duty cycle used in the modulation process, which has the structure shown in (2).

$$d_B(t) = d_{B1} \cos(\omega_1 t + \delta_1) - d_{B2} \cos(\omega_2 t + \delta_2) \quad (2)$$

Where $d_{B1}$ and $d_{B2}$ are the modulation indices associated with the expected voltage magnitudes in the internal voltage. $\delta_1$ and $\delta_2$ are the angular phases necessary to control the input and output of the active power in the converter.

The branch current $i_B$ is made up of two frequency components as well as the internal voltage. Its general structure is as presented in (3).

$$i_B(t) = \frac{I_1}{3}\cos(\omega_1 t + \theta_1) - \frac{I_2}{3}\cos(\omega_2 t + \theta_2) \quad (3)$$

In (3) it is noted that the magnitudes of the two components of the branch current correspond to a third of the currents of systems 1 and 2. This is because each phase of each system is connected to three branches, by what its current is divided equally under the assumption that the nine branches are perfectly balanced.

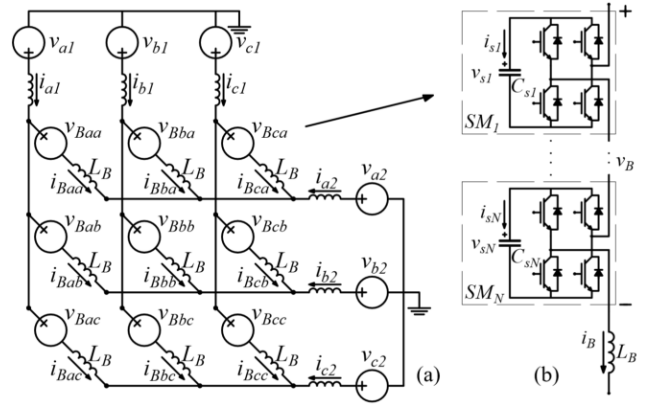

Figure 2. MMMC Topology

For its part, the current circulating in the storage capacitor is related to the current $i_B$ as follows:

$$i_s(t) = d_B(t) \times i_B(t). \quad (4)$$

### B. Back-to-Back Modular Multilevel Converter (BTB-MMC)

BTB-MMC changes the frequency through an intermediate stage in DC, allowing a complete decoupling between the two external networks.

As shown in *Figure 3a*, the BTB-MMC is made up of two MMCs that share a DC link. Each of these MMCs interacts with one of the external networks through the six branches that compose it. Each branch, as in the MMMC, is composed of a controlled voltage source ($v_B$) in series with an inductor ($L_B$) that fulfills the same function as that the one used in the MMMC. Unlike MMMC, submodules are typically Half-Bridges capable of bidirectional current flow. A schematic of the submodule string is shown in *Fig 3b*. As for the duty cycle, as shown in (5), it will be made up of a DC component and an AC component at the corresponding network frequency.

$$d_B(t) = d_{B(dc)} \pm d_{B(ac)} \cos(\omega_x t + \delta_x), x \in \{1,2\} \quad (5)$$

Consequently, the branch current ($i_B$) will also possess the same frequency components of the internal voltage. Analyzing the BTB-MMC topology in *Fig 3a*, it is clear that the $I_{dc}$ current is divided into three and the AC system current 1 or 2 is divided into two equal parts when entering/leaving the converter. Equation (6) condenses the results of this analysis.

$$i_B(t) = \frac{I_{dc}}{3} \mp \frac{I_x}{2} \cos(\omega_x t + \theta_x), x \in \{1,2\} \quad (6)$$

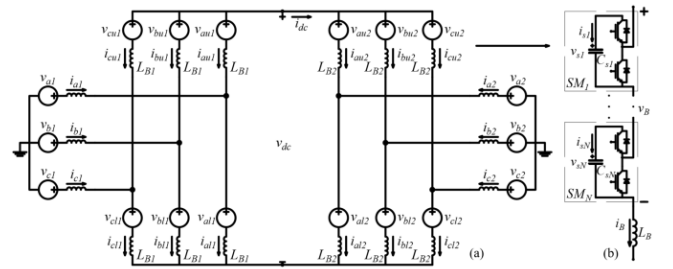

Figure 3. Back-to-Back MMC topology

## III. CONVERTER COMPARISON

In this section, design considerations regarding semiconductor requirements, reactive component requirements, operational losses, and fault tolerance are discussed. An appropriate analytical model is used for each converter to identify the key design variables. For the numerical comparison of the design variables, the two topologies considered will be designed based on a capacity of 450MVA/360MW, a line-to-line RMS voltage of 230kV, with a grid frequency ($f_1$) of 60Hz and a transmission frequency ($f_2$) of 20Hz. At the end of the section, the typical waveforms for each of the topologies studied are presented.

### A. Rated branch currents and branch inductor ($L_B$) selection

As evidenced in Section II, each of the topologies has a different branch current distribution. This is shown in *TABLE I*, where it is observed that the BTB-MMC topology is the one with the highest branch current. The reason this configuration has the highest branch current is because the component associated with system 1 (or 2) is divided into two paths between the upper and lower branches of the MMC. Meanwhile, the MMMC topology divides the currents into three of equal magnitude.

The branch inductor calculation can be calculated in several ways depending on the design criteria established. For example, one possibility is to size the inductor in order to reduce ripple in the current due to high frequency switching. An example of this approach is presented in [13], where an MMC is designed for application in HVDC. To calculate the inductor, the authors used a maximum ripple of 20%, calculated at the first harmonic component present, which is 40 times the frequency of the electrical network. On the other hand [14] it raises the selection of the inductor based on how the short-circuit current is handled on the DC-side. If the submodules are equipped with bypass-thyristors that protect them from short-circuit currents, the value of the inductor can be around 0.05pu. Otherwise, and in order to reduce the short-circuit current, the inductor must be selected with values between 0.10 and 0.15pu.

In this work, it is proposed to size the branch inductor as a function of the maximum voltage drop expected in it. Such voltage drop is represented by parameter $k$, which indicates the percentage of voltage drop in the branch inductor with respect to the nominal voltage of the external system to the converter. Considering a value of $k$ of 1.5% (2,817kV$_{peak}$), the inductance values shown in *TABLE II* are obtained.

The equations for each of the topologies were deduced based on the RMS voltage of the inductor. It is observed that the smallest inductor is the one obtained for the MMC connected to the side of $f_1$ ($x = 1$), with a value of 9.35mH, equivalent to 0.030pu. In contrast, its MMC counterpart connected to the side of $f_2$ ($x = 2$) has an inductor of 28.06mH (0.09pu). the ratio between the inductors is equivalent to the ratio between $f_1$ and $f_2$. In the case of MMMC it was obtained inductance of 13.31mH (0.0427pu). On the other hand, the nominal power is calculated as the product between the RMS voltage, the RMS current and the total number of inductors in the topology (which coincides with the number of branches).

TABLE I. RATED BRANCH CURRENT

|  | MMMC | BTB-MMC |
|---|---|---|
| $I_B$ peak (A) | 1064.96 | 1083.70 |
| $I_B$ rms (A) | 532.48 | 632.60 |
| $I_B$ avg (A) | 425.69 | 454.46 |

### B. Rated DC voltage and storage capacitor ($C_s$) selection

Regarding the design of the storage capacitor voltage ($v_s^\Sigma$), it must be considered that it consists of a component in DC ($V_s^\Sigma$) and several components in AC, which in (7) are condensed in a single term called ripple voltage ($v_{s_{ripple}}^\Sigma$).

$$v_s^\Sigma(t) = V_s^\Sigma + v_{s_{ripple}}^\Sigma(t) \qquad (7)$$

The DC component must be large enough to allow a correct synthesis of the internal voltages in each of the branches of the converter.

On the other hand, the AC components must be small enough to avoid unwanted oscillations in the capacitor voltage. As seen in (8) these components are directly proportional to the current flowing through the capacitor and inversely proportional to the value of capacitance.

$$v_{s_{ripple}}^\Sigma = \frac{1}{C_s^\Sigma} \int i_s dt \qquad (8)$$

In this context, the value of the capacitance is used as the main solution for the reduction of the ripple in the voltage. The higher the capacitance, the lower the ripple. However, large capacitance values imply higher costs, so a technical-economic balance is necessary.

*TABLE III* shows the DC voltage required for each of the topologies. It is clearly observed that all of them are in the same vicinity, with an average of 420.41kV. In the case of the MMMC this voltage will be the net of a branch at any instant of time, while in the case of the BTB-MMC this voltage corresponds to $V_{dc}$ in the DC stage.

The parameter $m$ is used in the equations, which is the maximum modulation index set at 0.9. This value of $m$ prevents the peak of the synthesized waves from reaching the DC voltage. In the equations, they also show the values $V_{B1}$ and $V_{B2}$, which represent the magnitudes of the internal voltages associated with each of the terminals of the converter.

TABLE II. BRANCH INDUCTOR AND RATED POWER

| Topology | Equation | $L_B$ (mH) | $S_{LB}$[a] (MVA) |
|---|---|---|---|
| MMMC | $L_B = \dfrac{kV_{m1,2}/\sqrt{2}}{\sqrt{\left(\frac{1}{\sqrt{2}}\frac{\omega_1 I_1}{3}\right)^2 + \left(\frac{1}{\sqrt{2}}\frac{\omega_2 I_2}{3}\right)^2}}$ | 13.31 | 9.55 |
| BTB-MMC | $L_{Bx} = \sum_{x=1}^{2} \dfrac{2kV_{mx}}{\omega_x I_x}, x \in \{1,2\}$ | 9.35+28.06 | 7.56x2 |

a. Total inductor power

TABLE III. RATED DC VOLTAGE

|  | MMMC | BTB-MMC |
|---|---|---|
| Equation | $V_s^\Sigma = \dfrac{V_{B1} + V_{B2}}{m}$ | $V_s^\Sigma = V_{dc} = \dfrac{2V_{B1,B2}}{m}$ |
| $V_s^\Sigma$ (kV$_{dc}$) | 419.71 | 421.11 |

To reduce the ripple voltage, it must be select the appropriate capacitance value. For this, through the algebraic manipulation of (4) and (8), the expressions shown in *TABLE IV* are obtained. These equations were derived by taking only the magnitudes of the frequency components of the ripple voltage. In the case of the MMMC, the ripple voltage has 4 components: $2\omega_1$, $2\omega_2$, $\omega_1 + \omega_2$ and $|\omega_1 - \omega_2|$. Meanwhile, in the BTB-MMC case, the ripple voltage has two components: $\omega_1$ and $2\omega_1$. Additionally, the parameter $k_r$ is defined, which represents the ripple as a percentage of the DC voltage of the capacitor. An acceptable value of $k_r$ is 10% as indicated in [14].

The capacitance values shown in *TABLE IV* represent the total capacitance of a branch. To determine the necessary capacitor in each submodule, the total capacitance must be multiplied by the total number of submodules. As in the case of branch inductors, the capacitor power is calculated as the product of the RMS voltage, the RMS current, and the total number of capacitors in the topology. Regarding the Energy-Power ratio, it is observed that the BTB-MMC is the one with the highest requirements among the two topologies.

### C. Power semiconductor requirements

According to the literature, IGBTs and IGCTs have been widely used in applications with modular converters. This widespread use lies in their current capabilities and blocking voltages. For high voltage applications, IGBTs can be achieved with blocking voltages up to 6.5kV and current capacity up to 750A. In the case of IGCTs they can be achieved with 10kV blocking voltage and current capacity of up to 1700A [15]. Regarding the switching frequencies, the IGBT can handle frequencies in the order of kilo-Hertz, while the IGCT has a limit of approximately 500Hz [13]. For comparative purposes and to avoid adding new variables to the study, the same semiconductor has been selected for the two topologies. The latter has been selected between since, for the considered topologies, commutation frequencies in the order of kilo-Hertz are not required [12]. The IGCT model chosen was the one that met the RMS current requirements in the branch inductor with the highest possible blocking voltage.

TABLE IV. STORAGE CAPACITOR, RATED POWER AND ENERGY-POWER RATIO

| Topology | Equation | $C_s^\Sigma$ (µF) | $S_{Cs}^a$ (MVA) | EP[b] (kJ/MVA) |
|---|---|---|---|---|
| MMMC | $C_s^\Sigma = \dfrac{I_{1,2}}{3k_r V_s^{\Sigma 2}}\left(\dfrac{V_{B1}}{2\omega_1} + \dfrac{V_{B2}}{2\omega_2} + \dfrac{2\omega_1(V_{B1}+V_{B2})}{\omega_1^2 - \omega_2^2}\right)$ | 98.39 | 1061.5 | 177 |
| BTB-MMC | $C_{sx}^\Sigma = \sum_{x=1}^{2} \dfrac{2}{k_r \omega_x V_s^{\Sigma 2}}\left(\dfrac{1}{2}\dfrac{I_x}{2}\dfrac{V_{Bx}}{2} + \dfrac{I_{xdc}}{3}V_{Bx} + \dfrac{I_x}{2}\dfrac{V_{dc}}{2}\right)$  $x \in \{1,2\}$ | 77.8+233.4 | 996.2 | 368 |

a. Total capacitor power. b. Total Energy-Power ratio.

*TABLE V* contains the most relevant information on the selected semiconductor. It is important to note that the required number of submodules per branch is 129. The equation used adds one more submodule as a reserve for reliability considerations.

TABLE V. SUMMARY OF SEMICONDUCTOR SELECTION

|  | MMMC | BTB-MMC |
|---|---|---|
| Device Type | IGCT | |
| Part Number | ABB 5SHX 19L6020 | |
| Blocking voltage – $V_{DRM}$ (V) | 5500 | |
| De-rated voltage – $V_{DC}$ (V) | 3300 | |
| Average on-state current – $I_{TAVM}$ (A) | 840 | |
| RMS on-state current – $I_{TRMS}$ (A) | 1320 | |
| On-state voltage – $V_{T0}$ (V) | 1.9 | |
| Slope resistance – $r_T$ (mΩ) | 0.9 | |
| Turn-on energy per pulse – $E_{on}$ (J) | 1.0[a.] | |
| Turn-off energy per pulse – $E_{off}$ (J) | 11.0[a.] | |
| (Diode) Turn-off energy – $E_{rec}$ (J) | 4.5[a.] | |
| Number of series submodules per branch/arm – N = ceil($V_s^\Sigma/V_{DC}$) + 1 | 129 | 129 Each MMC |
| Total number of devices | 4644 | 1588x2 |
| Semiconductor MVA (MVA) | 12863.88 | 8797.52 |

a. V$_{ref}$ = 3300V and I$_{ref}$ = 1800A

### D. Operating losses

The losses to be considered in the study are those produced in the semiconductor devices and in the reactive elements. Losses associated with auxiliary and support circuits will not be considered. *TABLE VI* presents each of the calculated loss components.

The losses in semiconductor devices can be subdivided into conduction losses and switching losses. Conduction losses result from the product between the average current and the voltage drop across the device during the conduction period. For the thyristor family (such as IGCT), the voltage drop is modeled with a minimum conduction voltage ($V_{T0}$) added to a resistive voltage drop defined by the resistor $r_T$ and the current $I_{on}$, which corresponds to the current flowing during the evaluated conduction period.

TABLE VI. OPERATING LOSSES

|  | MMMC | BTB-MMC |
|---|---|---|
| Conduction losses – $P_{on}$ (MW)  $P_{on} = (V_{T0} + r_T I_{on}) \times I_{B_{AVG}}$ | 5.650 | 2.023x2 |
| Switching losses – $P_{sw}$ (MW)  $P_{sw} = \left(\dfrac{E_{on}I_{sw}}{V_{on}^{ref}I_{on}^{ref}} + \dfrac{E_{off}I_{sw}}{V_{off}^{ref}I_{off}^{ref}} + \dfrac{E_{rec}I_{diode}}{V_{rec}^{ref}I_{rec}^{ref}}\right)\dfrac{V_s^\Sigma}{N}f_{sw}$ | 8.050 | 2.738x2 |
| Capacitor losses – $P_{Cs}$ (MW)  $P_{Cs} = \dfrac{0.5W}{1000VAR}S_{Cs}$ | 0.531 | 0.498x2 |
| Inductor losses – $P_{LB}$ (MW)  $P_{LB} = 0.02\% \times S_{conv}$ | 0.090 | 0.090 |
| Efficiency (%) | 96.17 | 97.14 |

Switching losses are characterized by being discrete amounts of energy that are lost due to switching events [13]. In the case of the IGCT, these losses are associated with the turn-on ($E_{on}$) and the turn-off ($E_{off}$) energies. In the case of the diode in antiparallel, its losses are associated with the reverse recovery energy ($E_{rec}$). To determine the total power lost by the commutation, the total lost energy must be multiplied by the commutation frequency ($f_{sw}$), which was selected as 3 times the grid frequency ($f_1$).

The reactive elements for the calculation of losses will be the branch inductor and the storage capacitor. Branch inductors are typically designed to minimize losses, so losses are not a big concern [16]. However, it is estimated that the losses due to them are around 0.02% of the total capacity of the converter [14]. For the capacitors used in this type of applications, the manufacturers specify an average loss of 0.5W/kVAR [17].

*E. Fault tolerance*

This section will analyze the ability of power circuits to handle/tolerate faults. Three types of faults will be specifically discussed, (i) converter faults, (ii) AC side faults and (iii) DC side faults.

The most common failure in this type of converters is the one that occurs in the submodules. Typically, each of them has a mechanical or solid-state switch connected in parallel, which acts in case of failure, disconnecting the faulted submodule. A common solution to compensate for the loss of the submodule removed from the string is to have reserve submodules that replace the failed submodule. Another solution is to redistribute the voltage between the healthy submodules by recalculating the voltage set-point of the storage capacitors [18], [19]. Some works such as [20], [21] present a more critical scenario in which an entire branch of an MMMC is rendered inoperative, either by electrical failure or by error in the control signals. Under this scenario, the reconfiguration of the MMMC to a Hexverter structure with reduced power and only six branches is proposed [22].

Regarding fault handling on the AC side, converters can be equipped with control loops for positive and negative sequence currents, making them capable of handling conditions in which the AC network is unbalanced [14].

Regarding DC faults, the only circuit susceptible to this type of fault is the BTB-MMC. For the MMC configuration with Half-Bridge submodules, the handling of fault currents in DC is not possible. When the fault occurs, the switches of each submodule are blocked in order to avoid damage to them, leaving the diodes in antiparallel working as an uncontrollable rectifier that feeds the fault from the AC side to the DC side. To overcome this problem, it can be used Full-Bridges instead of Half-Bridges on the submodules. Under this scenario, it is possible to zero the internal DC voltage of each branch in order to clear the fault [23].

*F. Waveforms*

Through simulations in EMTP-RV, the scenario where the converter operates at nominal capacity was simulated, sending 360MW from system 1 to 2 and injecting 270MVAR of reactive power from the converter to each of the systems. *Figure 4*, *Figure 5* and *Figure 6* show the expected waveforms.

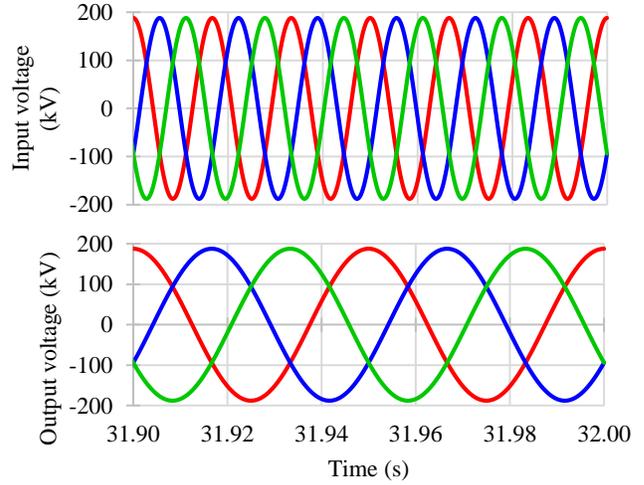

*Figure 4. Three-phase Input and Output voltages*

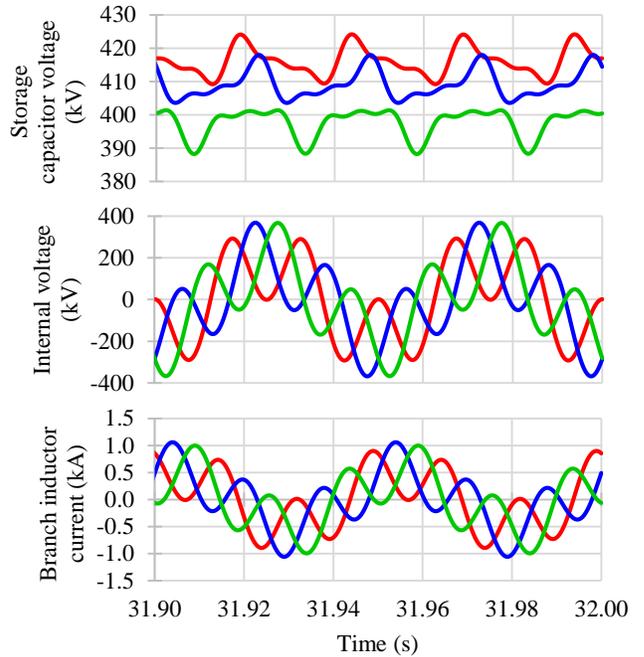

*Figure 5. Waveforms in MMMC*

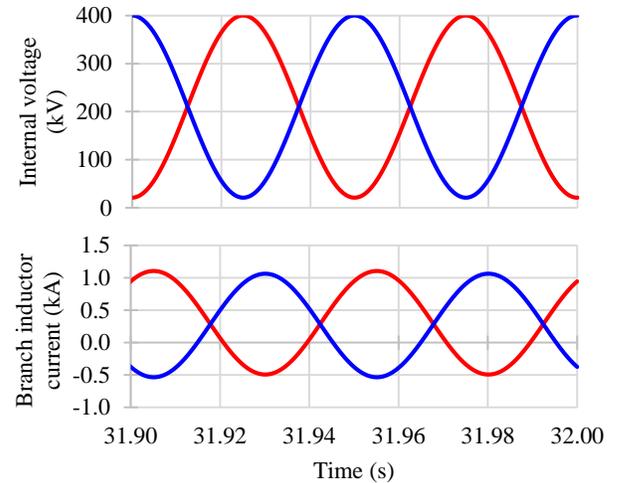

*Figure 6. waveforms in BTB-MMC*

In *Figure 5* is possible to observe the two frequency components both in the waves of the internal voltage and in the current of the branch inductor. The internal voltage peak does not exceed DC voltage, which is necessary for correct modulation. *Figure 6* also shows the DC and AC components in voltage and current. In the internal voltage it is observed that negative voltage values are never obtained, a necessary condition when using Half-Bridges in the MMC.

## IV. Conclusions

This article has presented a comparative evaluation of two candidate topologies to improve the transfer capacity of existing transmission lines by reducing the electrical frequency at which energy is transmitted. In general terms, since the topologies belong to the family of modular converters, some of the evaluated characteristics present similar results, such as the DC voltage of the storage capacitor and the number of submodules per branch. However, characteristics such as the rated branch current, the size of the branch inductor and storage capacitor, the Energy-Power ratio of the converter, the efficiency and the fault tolerance are significantly different. The BTB-MMC rated branch current is higher than the current in the MMMC. The BTB-MMC branch inductor on the 20Hz-Side is more than 200% higher than the corresponding on MMMC. The Energy-Power ratio of the BTB-MMC is 200% higher than the one in the MMMC. Finally, the fact that MMMC configurations are made up of Full-Bridges makes their fault tolerance much higher than its BTB-MMC counterpart, which is unable to handle failures on the DC side. With these significant differences it can be said that the MMMC configuration could be a better candidate for the benchmark application.

## Acknowledgment

The authors gratefully acknowledge support from Wisconsin Electric Machines and Power Electronics Consortium (WEMPEC), to the Fulbright-MinCiencias scholarship and to the professorial development scholarship of the Universidad del Norte. This work was supported by the New York Power Authority (NYPA), and the New York State Energy Research and Development Authority (NYSERDA).